\documentclass{aastex} 
\usepackage{emulateapj5} 
\usepackage{apjfonts,pstricks,amsmath,epsfig,pst-node}  

\makeatletter

\newenvironment{inlinefigure}{%
\def\@captype{figure}%
\noindent\begin{minipage}{0.999\linewidth}\begin{center}}
{\end{center}\end{minipage}\smallskip} 
\makeatother                                                   

\newcommand\ri{\mbox{$R\!-\!I$}}
\newcommand\vi{\mbox{$V\!-\!I$}}

\slugcomment{Accepted to {\it{The Astrophysical Journal}}}
\shortauthors{Gonzalez et al.}
\shorttitle{Tests of the LCDCS from EDisCS}
 
\begin{document}
\title{Tests of the Las Campanas Distant Cluster Survey
from Confirmation Observations for the ESO Distant Cluster
Survey\altaffilmark{1}}
\altaffiltext{1}{Based on observations obtained in visitor and service modes at 
the ESO Very Large Telescope (VLT) as part of the Large Programme 166.A--0162
(the ESO Distant Cluster Survey)}
  
\author{Anthony H. Gonzalez\altaffilmark{2,3}, Dennis Zaritsky\altaffilmark{4},
Luc Simard\altaffilmark{4,5}, Doug Clowe\altaffilmark{6}, 
Simon D. M. White\altaffilmark{7}}
   
\altaffiltext{2}{Harvard-Smithsonian Center for Astrophysics, 60 Garden Street,
MS-20, Cambridge, MA, 02138, USA}
\altaffiltext{3}{Current Address: Dept. of Astronomy, 
University of Florida, 211 Bryant Space Science Center, Gainesville, FL, 32611-2055, USA}
\altaffiltext{4}{Steward Observatory, University of Arizona, 933
   North Cherry Avenue, Tuscon, AZ 85721, USA}
\altaffiltext{5}{Current Address: Herzberg Institute of Astrophysics, National Research Council of Canada, 5071 W. Saanich Rd, Victoria, BC, V9E 2E7, Canada}
\altaffiltext{6}{Institut f\"ur Astrophysik und Extraterrestrische Forschung 
der Universit\"at Bonn, Auf dem H\"ugel 71, 53121 Bonn, Germany }
\altaffiltext{7}{Max-Plank Institut f\"ur Astrophysik, Karl Schwarzschild Strasse 1,
Garching bei Munchen, 85741, Germany}

\begin{abstract} 
                                                  
  The ESO Distant Cluster Survey (EDisCS) is a photometric and
  spectroscopic study of the galaxy cluster population at two epochs,
  $z \simeq 0.5$ and $z\simeq 0.8$, drawn from the Las Campanas
  Distant Cluster Survey (LCDCS).  We report results from the initial
  candidate confirmation stage of the program and use these results to
  probe the properties of the LCDCS. Of the 30 candidates targeted,
  we find statistically significant overdensities of red galaxies near
  28.  Of the ten additional candidates serendipitously observed
  within the fields of the targeted 30, we detect red galaxy
  overdensities near six.  We test the robustness of the published
  LCDCS estimated redshifts to misidentification of the brighest
  cluster galaxy (BCG) in the survey data, and measure the spatial
  alignment of the published cluster coordinates, the peak red galaxy
  overdensity, and the brightest cluster galaxy. We conclude that for
  LCDCS clusters out to $z \sim 0.8$ , 1) the LCDCS coordinates agree
  with the centroid of the red galaxy overdensity to within 25\arcsec
  ($\sim 150 h^{-1}$ kpc) for 34 out of 37 candidates with $3\sigma$
  galaxy overdensities, 2) BCGs are typically coincident with the
  centroid of the red galaxy population to within a projected
  separation of 200$h^{-1}$ kpc (32 out of 34 confirmed candidates),
  3) the red galaxy population is strongly concentrated, 
  and 4) the misidentification of the
  BCG in the LCDCS causes a redshift error $>$ 0.1 in 15-20\% of the
  LCDCS candidates.  These findings together help explain the success
  of the surface brightness fluctuations detection method.

\end{abstract}

\section{Introduction}

Observations of the distant galaxy cluster population are being driven
by a new generation of catalogs that provide statistically significant
samples of hundreds to thousands of candidate clusters between
redshifts 0.5 and 1.  These catalogs can either serve as large
statistical samples without appealing to any further observations, for
example to measure the cluster-cluster correlation function
\citep{gon2002}, or as input for follow-up studies of selected
subsamples.  The European Southern Observatory Distant Cluster Survey
(EDisCS) is a detailed follow-up study of 20 clusters, 10 at $z \simeq
0.5$ and 10 at $z \simeq 0.8$, drawn from the 1073 candidate clusters
cataloged by the Las Campanas Distant Cluster Survey
\citep[LCDCS;][]{thesis,gon2001}.  This paper describes results from
the preliminary effort to confirm the set of cluster candidates that
will be the focus of the more extensive observations of the EDisCS.

The value of a cluster catalog is greatly enhanced for any application
if the catalog includes measurements of the redshift and mass of each
candidate cluster.  Due to the size of recent catalogs, it is
impractical to obtain {\it spectroscopic} redshifts or masses for a
significant fraction of the catalog. Most catalogs now provide an
estimate of these properties drawn solely from the survey data
\citep[see][]{post96,gon2001}.  Superior survey data, for example
deeper images or multiple colors, should improve the reliability of
the estimated parameters, but they decrease the observing efficiency.
The optimum balance between the fidelity of the cluster catalog and
observational efficiency is not evident, and will depend on the
scientific aims. \citet{gon2001} provided a catalog from what is
arguably the most observationally efficient method (10 nights at a 1m
telescope produced a catalog of $\sim$ 1000 cluster candidates out to
$z \sim 1$ over an area covering 130 sq. degrees), but which might in
turn provide the least robust estimates of the cluster redshifts and
masses, and which, even more importantly for some potential uses of
the catalogs, may include a larger fraction of false detections. Using
observations in multiple filters that are $\sim$ 3 magnitudes deeper
than original survey data, we examine whether the false positive rate
quoted originally for the LCDCS is valid and whether the LCDCS cluster
coordinates and estimated redshifts are confirmed using deep,
multicolor data.

The LCDCS generated a catalog of concentrations of photons on the sky
rather than galaxies \citep{dal1995,zar1997,gon2001}. Significant
fluctuations in the background sky are classified into various
categories, including high redshift clusters.  The cluster redshift is
estimated using the magnitude of the brightest galaxy near the surface
brightness fluctuation, which is presumed to be the brightest cluster
galaxy (BCG).  The redshift-magnitude relationship for BCGs is
calibrated using spectroscopy of a sample of $\sim$ 20 clusters.  The
cluster mass is estimated using the peak brightness of the convolved
surface brightness map, calibrated using a sample of $\sim 10$
clusters with X-ray temperature and velocity dispersion measurements.
The uncertainties in each of these estimators, and of the false
positive and negative rates, are discussed by \citet{gon2001} in their
presentation of the LCDCS.  Our examination of these issues here
utilizes multifilter images of 30 targeted fields, which contain 40
LCDCS candidate clusters, obtained with ESO's Very Large Telescope
(VLT) as part of the initial stage of the EDisCS.

The format of this paper is as follows. In \S 2 we present the data
utilized in this analysis and details of the reduction procedure.  We
then examine the two-color photometry in \S 3 to confirm or reject the
cluster candidates observed by EDisCS and test whether the fractional
contamination is consistent with that given by \citet{gon2001}. In \S
4 we test the robustness of the estimated redshifts quoted for the
LCDCS, which are based upon the magnitude of the brightest cluster
galaxy, with particular emphasis on the potential problem of BCG
misidentification.  In this section we also quantify the offsets
between the LCDCS coordinates, the locations of the brightest cluster
galaxies, and the peak of the projected galaxy distribution. Next, we
briefly comment upon the LCDCS mass estimates in \S 5 and compare the
LCDCS surface brightness with other observable quantities.  Finally, a
summary of the results and brief discussion of forthcoming work are
presented in \S 5.  For all physical distances in this paper we assume
a flat, $\Lambda$CDM cosmology with $\Omega_0$=0.3.

\section{Data}

\subsection{Sample Selection}

The data presented here are part of the ESO Distant Cluster Survey, an
ongoing ESO large program to examine a set of $\sim$ 10 massive
clusters in each of two distinct redshifts regimes, $z\simeq0.5$ and
$z\simeq0.8$. The cluster candidates for this study are drawn from the
Las Campanas Distant Cluster Survey (Gonzalez et al.  2001), with
candidate selection constrained by the published redshift estimates
and surface brightnesses corrected for Galactic absorption,
$\Sigma_{cor}$.  Candidates are selected to be among the highest
surface brightness detections at each redshift in an attempt to
recover some of the most massive clusters at each epoch. Because some
of the brightest detections, especially at higher estimated redshifts,
turn out to be spurious (for example, scattered light or tidal
material around nearby galaxies) and because there is a factor of two
scatter in the relationship between $\Sigma_{cor}$ and mass (as
measured from $T_X$ or $L_X$; \citealt{gon2001}), we visually classify
all of the candidates that satisfy our initial criteria (RA
constraints, $\Sigma_{cor} > 8\times10^{-3}$ counts s$^{-1}$
arcsec$^{-2}$, $0.45 \le z \le 0.55$ or $0.75 \le z \le 0.85$) to
select the most probable, massive clusters. The distribution in
$\Sigma_{cor}$ of the observed list of candidates (not all visually
approved candidates were observed) is shown in Figure
\ref{fig:sbdist}.  A second set of criteria, based on the data
presented here, are applied to select the final 20 clusters that will
be the focus of the EDisCS and will be discussed in the presentation
of the deeper photometry of the EDisCS clusters \citep{white02}.  The
final sample represents our best effort to select a subsample of the
most massive clusters at the two epochs, but is neither complete
within the survey volume nor necessarily unbiased with respect to the
LCDCS catalog.

\begin{figure*}
\epsscale{0.5}
\plotone{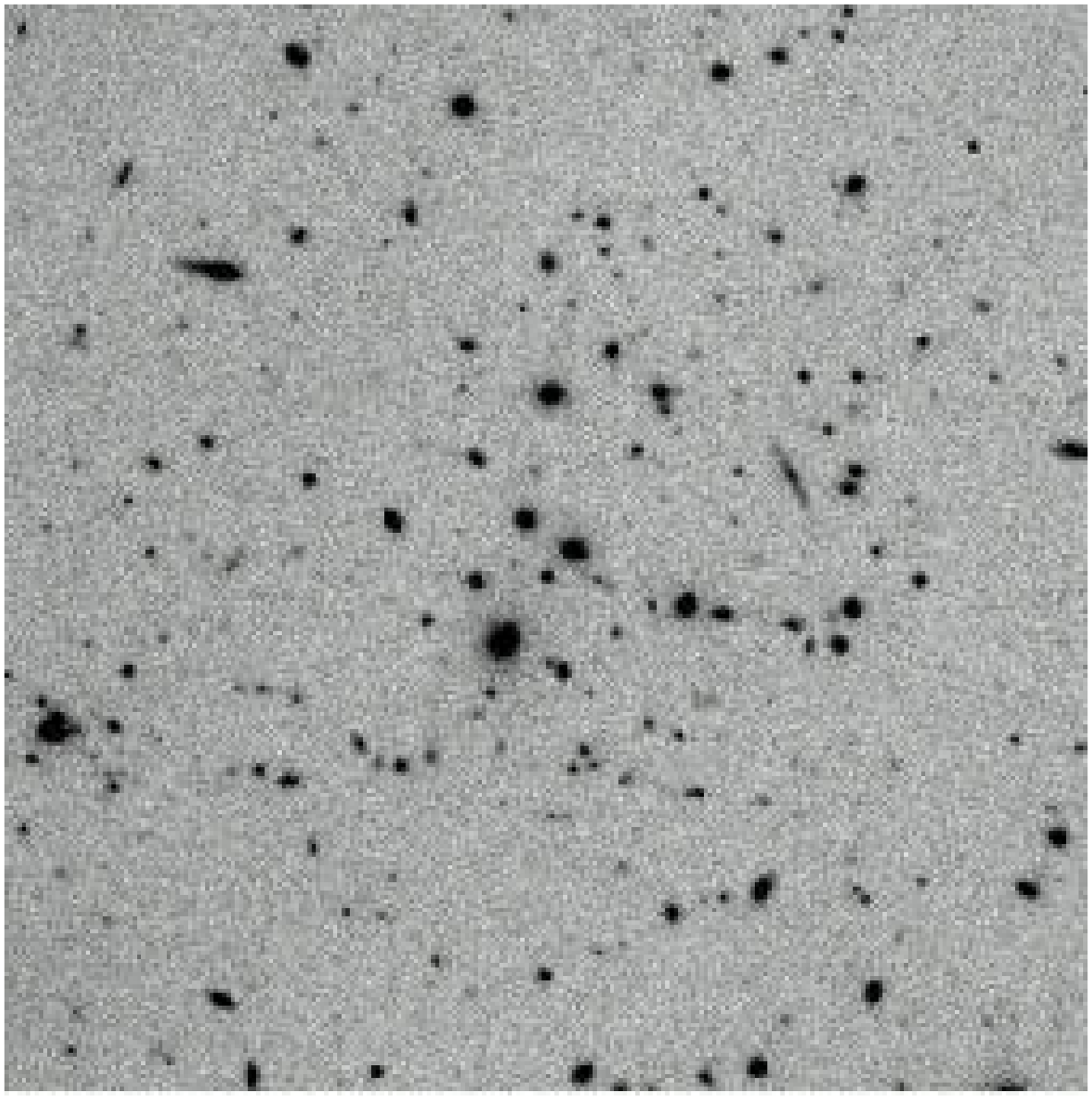}\hskip 0.1cm \plotone{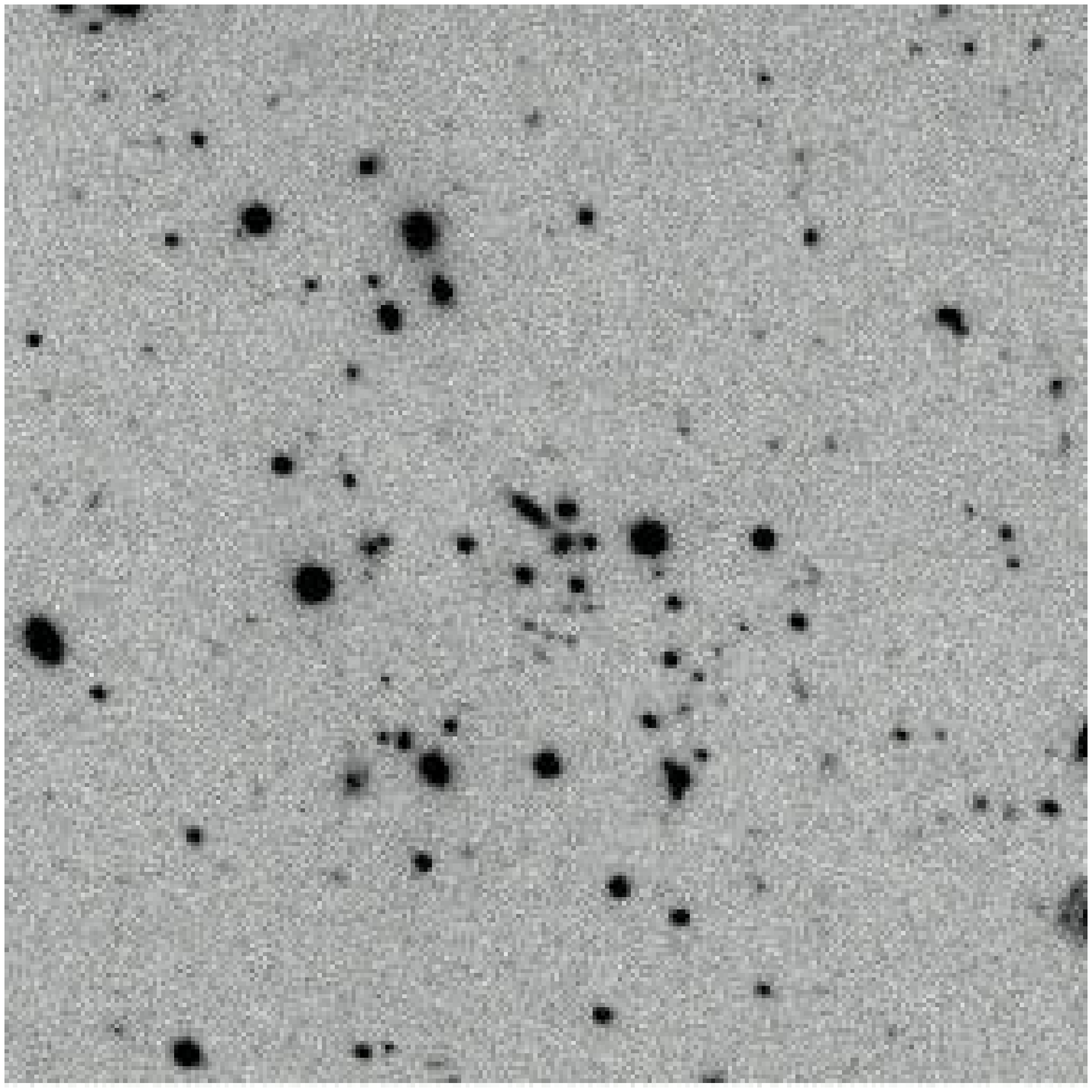}
\hskip 0.15cm \plotone{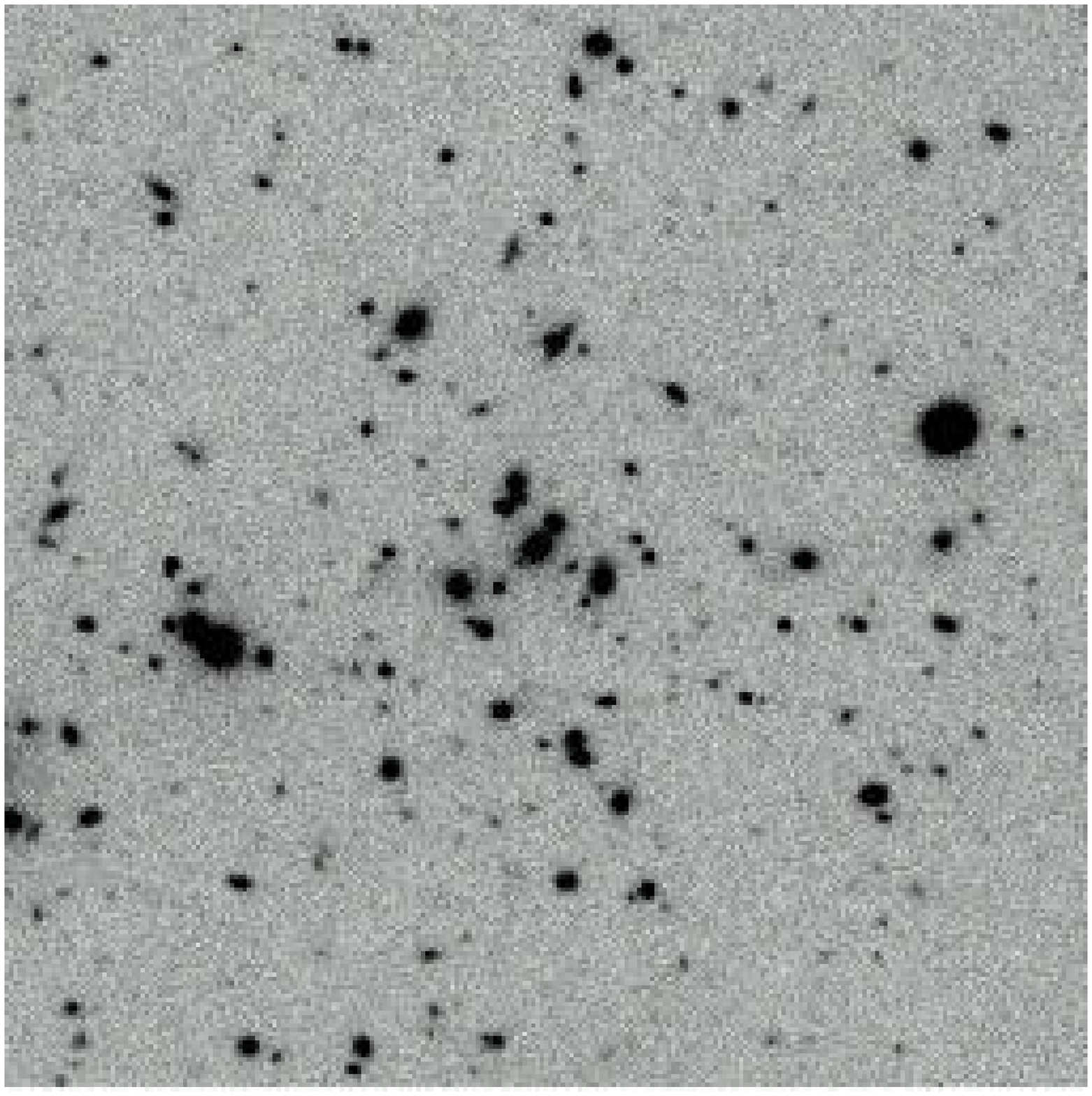}\hskip 0.1cm \plotone{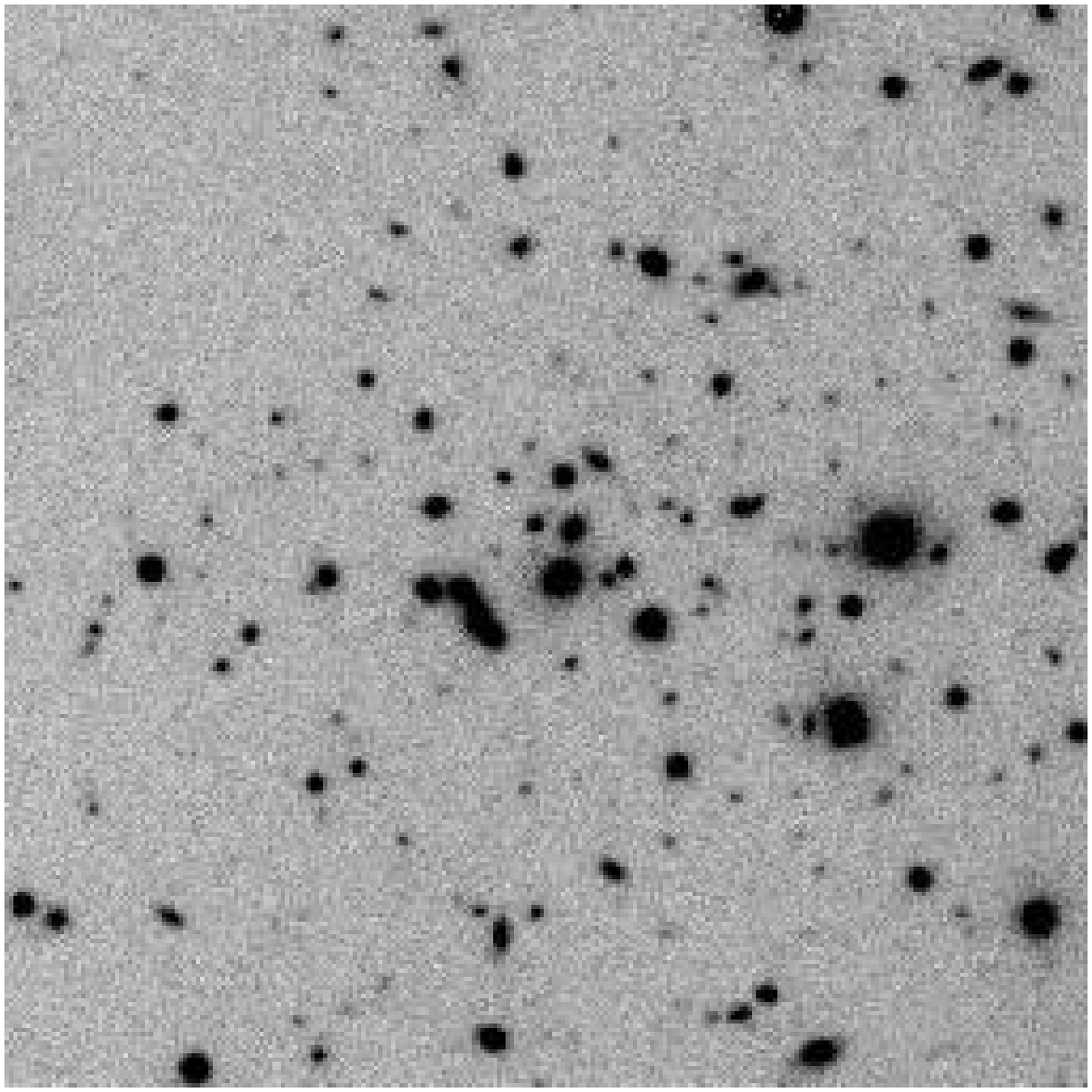}
\epsscale{1.0}
\figcaption{$I-$band images of four of the cluster candidates targeted by
this program. The field of view for each image is $2\arcmin\times2\arcmin$.
The two candidates on the left, LCDCS 0713 (far left) and LCDCS  0188, have estimated
redshifts $z\sim0.5$, while those on the right, LCDCS 0172  and LCDCS 0531 (far
right),
have estimate redshifts $z\sim0.8$. These candidates are neither the best
or worst in our sample, with galaxy overdensities detected at the 5-6 $\sigma$
significance level (see \S \ref{sec:clusterconf}).
\label{fig:images}}
\end{figure*}

Using the estimated contamination rate for the LCDCS from
\citet{gon2001} of $\sim$ 30\%, we targeted thirty candidates in an
effort to obtain twenty confirmed clusters.  We initially observed 11
at $z\simeq0.5$ and 13 at $z \simeq0.8$; the final 6 candidates were
selected to replace false detections, poor systems, and clusters with
galaxy colors inconsistent with the desired redshift intervals. Images
of four candidates, two at each epoch, are shown in Figure \ref{fig:images}.
These
thirty fields also serendipitously contain ten additional cluster
candidates from the LCDCS catalog, yielding imaging for a total of 40
LCDCS candidates. The ten serendipitous clusters are a more
representative selection from the LCDCS catalog, although some may be
at the same redshift as the target cluster and therefore be part of a
larger association of clusters and groups.

\begin{inlinefigure}
\plotone{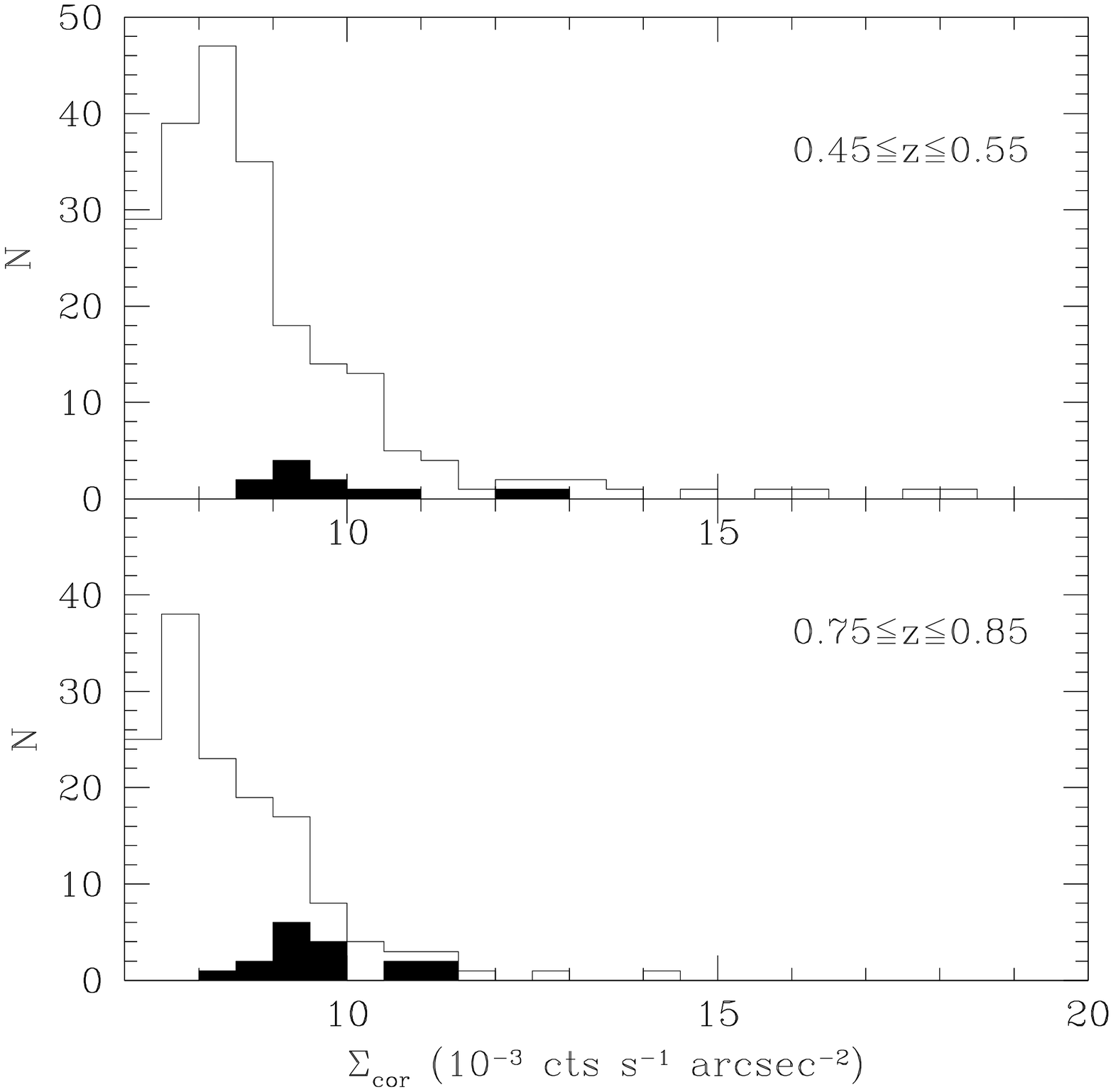}
\figcaption{Distribution of peak surface brightness from the
LCDCS catalog for the two epochs probed by EDisCS. The filled
histograms correspond to the primary EDisCS targets, while the open
histograms correspond to all clusters in the LCDCS catalog at
these epochs. \label{fig:sbdist}}
\end{inlinefigure}

\subsection{Observations and Image Analysis}

Preliminary imaging for the EDisCS was carried out in service mode
during January and February of 2001 and in visitor mode on March 19,
2001 using the FOcal Reducer/low dispersion Spectrograph 2 (FORS2) in
direct imaging mode on the UT2/Kueyen telescope on Paranal, Chile. The
field of view is $6.8\arcmin\times6.8\arcmin$, with corresponds to a 
physical extent of $\sim1.7-2.2 h^{-1}$ Mpc for the clusters in our sample.
In total, thirteen $z\sim 0.5$ candidate fields and seventeen $z\sim 0.8$
candidate fields were observed for 20 minutes in each of two
passbands: $I_\mathrm{B}$ and $R_\mathrm{Sp}$ for the $z\sim 0.8$
sample and $I_\mathrm{B}$ and $V_\mathrm{B}$ for the $z\sim 0.5$
sample.  In each passband, the observing time was split into 4
exposures of 5 minutes each, dithered by $\sim 10\arcsec$ between
exposures.  The service observing was only performed in photometric
conditions with seeing better than $1\arcsec$ seeing.  The conditions
on March 19 were $\sim 0\farcs8$ seeing and thin cirrus.  Comparison
of the images for the three clusters observed on March 19, which were
selected for the EDisCS final sample, to images taken later in
photometric conditions shows that the extinction varied between 0 and
$0.15$ magnitudes during the night.
 
The images had the bias removed by first subtracting a medianed master
bias frame, taken each night, and then performing and subtracting a
linear fit to the overscan strip residuals.  Because the data was
taken in 4-port readout mode, this step was done separately for each
quadrant.  Medianed nightsky flats were then constructed for each
week's data and used to flatfield the images.  Our visual inspection
of the flatfielded data did not find any evidence for dust spots or
other large-scale fluctuations, suggesting that the flatfield did not
change over the course of a week.  The images were then aligned using
integer pixel offsets and averaged with $3\sigma$ clipping to remove
cosmic rays.
 
For each cluster candidate field, the images taken in the two filters
were carefully registered (when needed) using IRAF/IMSHIFT. Galaxy
aperture photometry was performed using SExtractor version 2.2.1
\citep{ber96}.  SExtractor was used in ``two-image" mode using the
I$_B$ image as the detection image.  The detection and analysis
thresholds were set to 1 $\sigma_{bkg}$ with a minimum area of 5
pixels and a minimum contrast parameter for deblending of 0.001. To
maximize the signal-to-noise of our measured colors, we used an
aperture diameter of $2\arcsec$.  Instrumental aperture magnitudes
were calibrated using the airmasses of our observations and the FORS2
photometric zeropoints produced by the Quality Control Group at
ESO/Garching for the nights of our observations.

\section{Cluster Confirmation}\label{sec:clusterconf}

The first goal of the EDisCS program is to confirm a subset of the
candidate clusters selected from the LCDCS.  As described by
\citet{nelson2001}, photometric data alone provides various ways in
which to identify a cluster.  Clusters are by definition a localized
excess of galaxies, but these can be difficult to identify when
superposed on a sea of foreground and background galaxies. One
approach at increasing the contrast between the cluster and
contaminating field has been to utilize the observational result that
clusters contain galaxies that appear to be nearly as red (evolved) as
possible at each epoch out to at least $z \sim 1$
\citep{stanford1998}.  By examining the clustering of these red
galaxies, which lie along the passively evolving E/S0 sequence, the
clusters are more easily identified. Our approach is quantitatively
similar to that of \citet{olsen2001}, who divide their galaxy catalogs
into color slices and then employ an adaptive smoothing kernel to
detect statistically significant galaxy overdensities, and
qualitatively similar to the approach demonstrated by
\citet{glad2000}.  Like \citet{olsen2001}, we use adaptive kernel
smoothing, starting from an Epanechnikov kernel with $h=30\arcsec$
\citep[see][for more details on adaptive smoothing]{mer94}.  We
restrict the magnitude range of the galaxies used for the maps on the
basis of the estimated redshifts of the cluster candidates. For
candidates with estimated redshifts $z_{est}<0.7$ we include galaxies
with $18<m_I<21.5$, while for higher redshift candidates we raise the
limiting magnitude to $m_I$=23. We refrain from any further refinement
of the magnitude range to match individual clusters because the
accuracy of the redshift estimates is not yet firmly established.

Rather than using a fixed color slice set by the cluster's estimated
redshift, we search color space to optimize the contrast between the
cluster signal and the galaxy background. We step through color in
increments of 0.05 mag, deriving an adaptively smoothed density map at
each step using galaxies with colors within 0.5 mag of the central
color. This approach neglects the slope of the color-magnitude
relation for E/S0 galaxies, but the color change due to the slope is 
significantly less than 0.5 mag \citep{stanford1998}.  We define the optimal 
color to be the one that yields the highest significance galaxy overdensity within
1$\arcmin$ of the coordinates given in the LCDCS catalog.

\begin{figure*}
\plotone{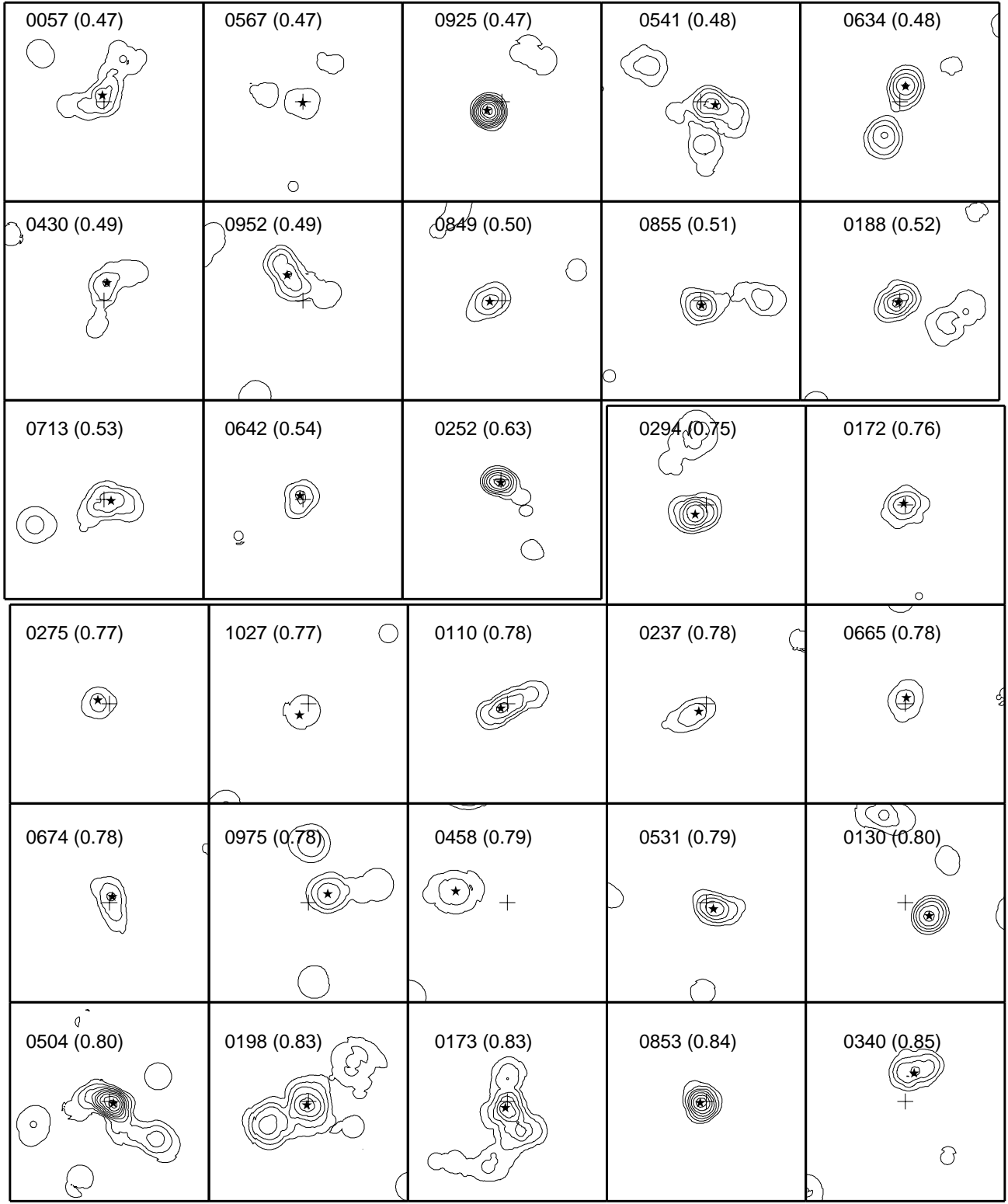}
\figcaption{Galaxy overdensity maps for each of the primary
EDisCS candidates. North is up and East is to the left. The labels
in each panel are the identification number and estimated redshift from
the LCDCS catalog, and the field of  
view for each panel is 
160$\arcsec\times$160$\arcsec$, centered on the coordinates from
\citet[][cross symbols]{gon2001}. Contours are separated by $1-\sigma$
intervals, with the lowest contours corresponding to a $2-\sigma$
threshold level. The stars correspond to the peak galaxy
overdensity in each field. Candidates with $z_{est}<0.7$ (top) were imaged
in $\ri$; those with $z_{est}\ge0.7$ (bottom) were imaged in $\vi$. We detect 
$>3-\sigma$ excesses within 1$\arcmin$ of all thirty EDisCS candidates. 
\label{fig:contour1}}
\end{figure*}

\begin{figure*}
\plotone{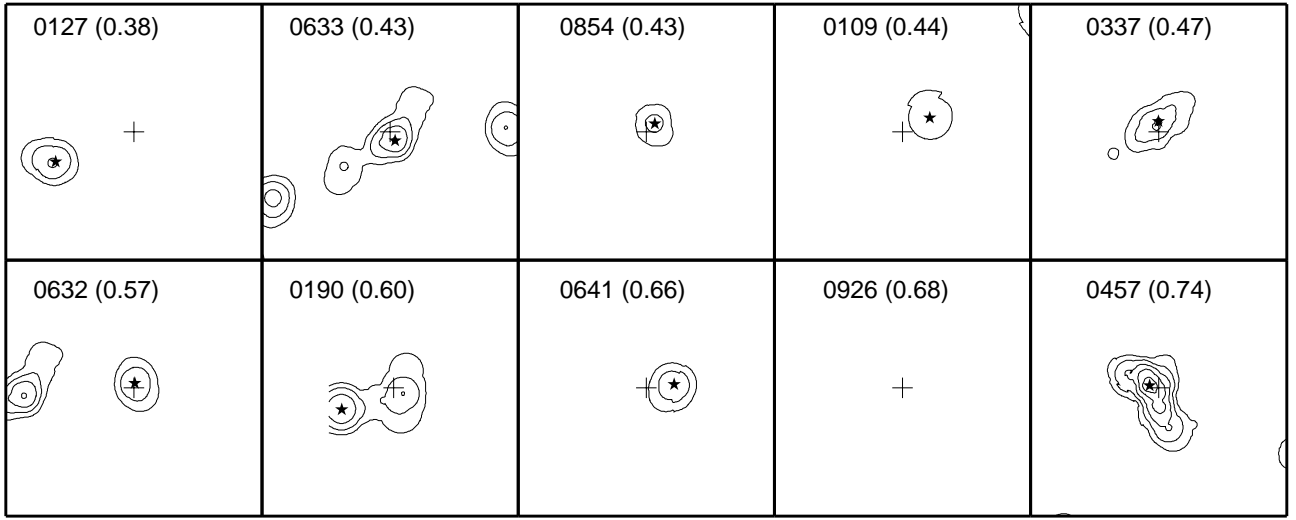}
\figcaption{Galaxy overdensity maps for the serendipitously imaged
LCDCS candidates. The field of view, labels, contours, and symbols are the same
as in Figure \ref{fig:contour1}. Statistically significant (3$\sigma$) galaxy overdensities
are found within 1$\arcmin$ of  eight  of the ten candidates. For LCDCS 0190,
the contours are truncated by the images boundary.
\label{fig:contour2}}
\end{figure*}

The results of the adaptive kernel smoothing are shown in Figures
\ref{fig:contour1}--\ref{fig:contour2} and Table
\ref{tab:confirmation}. We estimate the statistical significance of
galaxy overdensities by constructing a histogram of the pixel
distribution in the smooth maps and making the approximation that
84.15\% of pixels in the density map must exceed the mean ($\simeq0$)
by less than 1$\sigma$.  To assess the likelihood that detected
overdensities are associated with the original LCDCS detections, we
also compute the probability that an overdensity will lie within a
given distance of a surface brightness detection purely by chance.
Based upon the number of detected $3\sigma$ ($4\sigma$) overdensities,
we calculate that there is a 6\% (2\%) probability that a random
overdensity peak will lie within $25\arcsec$, or 2.5 times the scale
length of the smoothing kernel employed in the LCDCS. For a 1$\arcmin$
radius, the probability is 33\% (11\%).  Thus, detections within
25$\arcsec$ are unlikely to be chance superpositions at roughly the
3$\sigma$ confidence level, while those within 1$\arcmin$ are
significant at roughly the 1$\sigma$ level.

\epsscale{2.0}

\epsscale{1.0}

For the targeted EDisCS sample, 29 out of 30 candidates have a
$3\sigma$ peak present within 1$\arcmin$, with the one remaining
detection having a significance of 2.9$\sigma$. Of these 29, in only
one instance is the separation between the peak galaxy overdensity and
the original coordinates greater than 25$\arcsec$.  If we
conservatively consider both this detection and the 2.9$\sigma$
detection to be failed candidates, then the success rate for the
targeted EDisCS candidates is 93\%.  We conclude that the LCDCS
catalog at these values of $\Sigma_{corr}$ even out to $z_{est} \sim
0.8$ has a low false positive rate when augmented by visual
inspection.

As expected, the success rate is lower for the serendipitously
observed LCDCS clusters because these candidates lack the surface
brightness selection criteria and visual culling applied to the
primary EDisCS targets. Based upon the redshift distribution of these
ten candidates and the contamination estimate published by
\citet{gon2001}, we predict that 4$\pm$2 of these candidates should be
spurious. In practice, we find $3\sigma$ peaks within 1$\arcmin$ of 
8 out of 10 candidates.  The separation between the peak galaxy
overdensity is greater than 25$\arcsec$ for two of these detections.
If we again assume that these detections are not associated with the
LCDCS detections, then we derive a conservative bound on the success
rate of 60\%, which is consistent with the predicted rate.
\begin{figure*}
\epsscale{1.70}
\vskip -3.7cm
\plotone{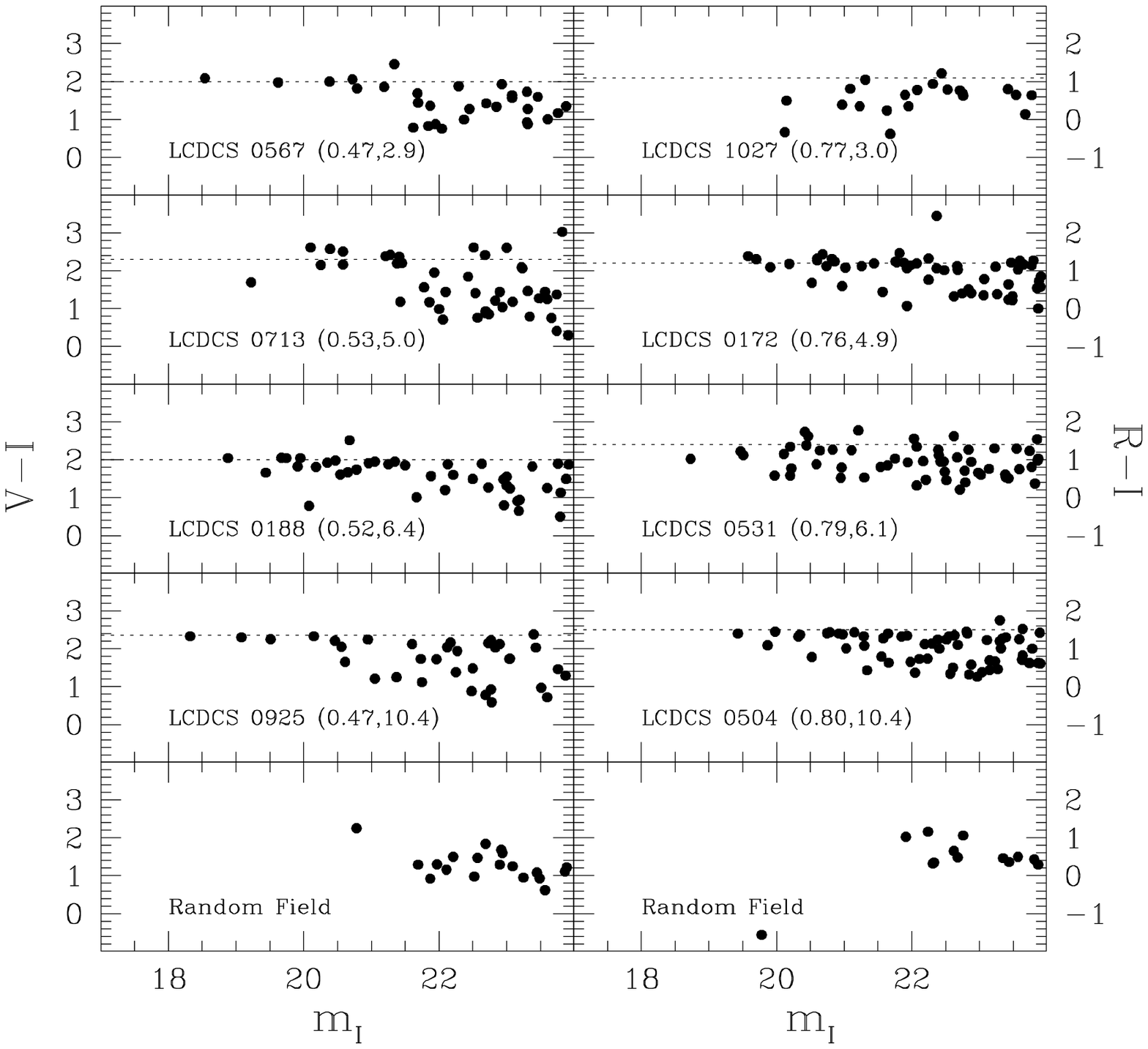}
\vskip -1.0cm
\figcaption{Color-magnitude diagrams for a subset of the cluster
candidates targeted by the ESO Distant Cluster Survey, with two random fields
shown for comparison. The plotted clusters are selected to span the range of
redshifts and galaxy overdensities covered by the full EDisCS sample. For the lower redshift
clusters (left), the color is $\vi$; for the higher redshift clusters (right)
$\ri$. The data points plotted in each panel include all galaxies within 
30$\arcsec$ of the published LCDCS coordinates. The dashed lines denote the 
color that yields the highest significance overdensity in the adaptively 
smoothed galaxy density maps. The numbers listed next to the cluster names
are the estimated redshift and the statistical significance of the detected
galaxy overdensity. 
\label{fig:cmd}}
\epsscale{1.0}
\end{figure*}

While the significant overdensities are indicative of clusters, we
must still confirm the LCDCS redshift and mass estimates. The best fit
colors from our adaptive smoothing procedure provide a first crude
test of the estimated redshifts.  Because the candidates are intended
to fall into two redshift bins, the colors of the clusters should be
consistent within each redshift bin and one might expect them to be
consistent with the colors of evolved galaxies at the two redshift
epochs. To refine our cluster color estimates, we rerun the adaptive
kernel algorithm using a narrower color window ($\pm$0.2 mag) than was
used for initial detection. In Figure \ref{fig:cmd} we plot
color-magnitude diagrams for a subset of our cluster candidates, overlaid
with dashed lines to denote the optimal color that we derive using the
adaptive kernel. The 
plotted candidates are selected to span the range of redshifts and galaxy 
overdensities covered by the full EDisCS sample.
In Figure
\ref{fig:colev} we plot this color ($\vi$ and $\ri$ for the low and
high redshift candidates, respectively) vs. $z_{est}$, and overplot
model tracks for an evolved single burst stellar population
\citep{bru93,cha96}. We caution that our colors cannot be
straightforwardly converted into redshift estimates (as one could
attempt to do from this Figure) because our method provides only a
crude color for the E/S0 sequence. However, a rough comparison to the
expected colors for evolved stellar populations suggests that the
reddest clusters do have colors consistent with their $z_{est}$, while
some other clusters appear to be significantly bluer than expected (up
to $\sim$0.6 mag), particularly among the $z\sim 0.8$ sample.  Without
spectroscopy we cannot reach a definite interpretation of this result,
but we can enumerate possible explanations. The scatter could be partially 
driven by the inclusion of overdensities that are alignments of poorer groups
along the line of sight rather than single clusters. Such detections
would have a broad color range rather than a well-defined red
sequence. 
Similarly,
rich foreground clusters could also contaminate the observed color-magnitude
diagrams of the targeted candidates.  
Chance alignments are unlikely in the LCDCS due to the small
size of the detection kernel \citep{gon2001}; however, such alignments
cannot presently be discounted in the galaxy overdensity maps. 
We anticipate that these effects are minor, but confirming this 
expectation requires more data.

Two additional, more interesting possibilities are that
1) the clusters have a well-developed
E/S0 sequence and are at lower redshifts than originally estimated, or
2) the clusters do not have a well-developed sequence and our color
matching finds a higher density contrast between cluster and field at
colors that are intrinsically bluer than the E/S0 sequence.  The
former conclusion would pose a serious problem for the LCDCS catalog
because many high redshift cluster candidates could turn out to be at
lower redshifts (by as much as 0.3 to 0.4); the latter conclusion
would have interesting implications both for models of the evolution
of cluster galaxies between $z=1$ and 0.5, and for selection of
clusters on the basis of the red sequence population
\citep[see][]{glad2000}. Our ongoing deeper photometry and spectroscopy of these
clusters will provide the necessary data to discriminate between all these
possibilities.

In concluding this section, we note that if the first possibility
(systematically underestimated redshifts) is correct, then there are
several factors that could contribute to such an error, including (1)
misidentification of the BCG when estimating the cluster redshift, (2)
failure of the BCG magnitude-redshift relation due to physical
evolution of the BCGs, or (3) failure of the BCG magnitude-redshift
relation in \citet{gon2001} at $z\approx0.8$ due to insufficient of
calibration data at this epoch.  We examine the possibility of BCG
misidentification in \S 4, finding that this factor alone is likely
insufficent to explain the range of observed cluster colors.  Physical
evolution is also an unlikely culprit, as various studies of BCGs out
to $z\sim1$
\noindent \citep{ara1998,burke2000,n02} find mild or no luminosity
evolution. If the redshifts are systematically underestimated, the
most probable culprit is failure of the magnitude-redshift relation
due to insufficient calibration data at this epoch - only one cluster
in the calibration set has a spectroscopic redshift at $z>0.7$.

\begin{inlinefigure}
\plotone{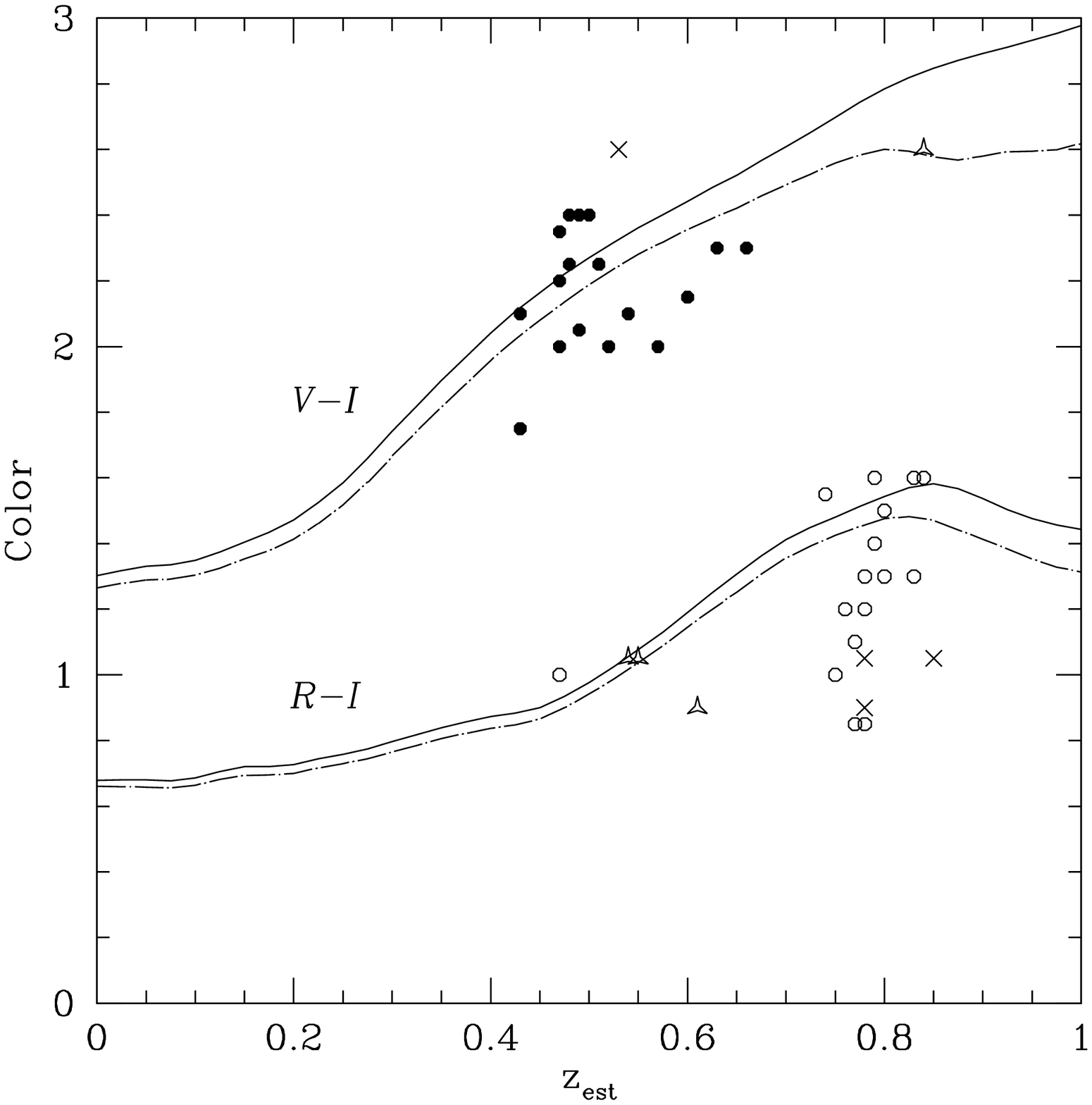}
\figcaption{Color of detected galaxy overdensity as a function of 
published
estimated redshift (see \S \ref{sec:zest}). Clusters with $\ri$ data are denoted by open
circles; those with $\vi$ data are denoted by filled circles. Overlaid are
passive evolution models for a flat
$\Lambda$CDM cosmology ($\Omega_0$=0.3). The dashed and solid lines are the
formation redshifts of 2 and 5, respectively. In the cases where BCG misidentification
leads us to revise the redshift estimate by more than 0.15 (see \S \ref{sec:bcg}), we
use crosses and open triangles to denote the original and revised redshift estimates, respectively.
\label{fig:colev}}
\end{inlinefigure}

\section{Brightest Cluster Galaxies}
\label{sec:bcg}

The estimated redshifts of each LCDCS cluster are derived using the
BCG magnitude-redshift relation, which has small scatter out to
$z\simeq1$ once a correction for cluster ``richness" is applied
\citep[e.g.][]{san1988,ara1993,ara1998,col1998} and should yield
redshift estimates accurate to $\pm$17\% \citep{post95}.  The key
concern with this approach, particularly for the LCDCS data, which are
shallow and do not include colors, is the possibility that the BCG has
been mis-identified.  We use the VLT data to address two aspects of
this concern. First, we use the color data to improve our BCG
selection \citep[see][]{n02}, and quantify any changes that occur in
our redshift estimates due to the potential selection of a different
galaxy as the BCG.  Second, we use the deeper imaging to estimate the
relative contributions of the cluster galaxies and the BCG envelope to
the surface brightness feature originally used to identify the
cluster.  \citet{gon2000} note that the BCG in Abell 1651, a nearby
cluster, contributes $\sim 36$\% to the total cluster light, so we ask
whether the LCDCS coordinates are more closely aligned with the BCG or
the peak of the galaxy distribution in the smoothed galaxy density
maps.

\begin{inlinefigure}
\plotone{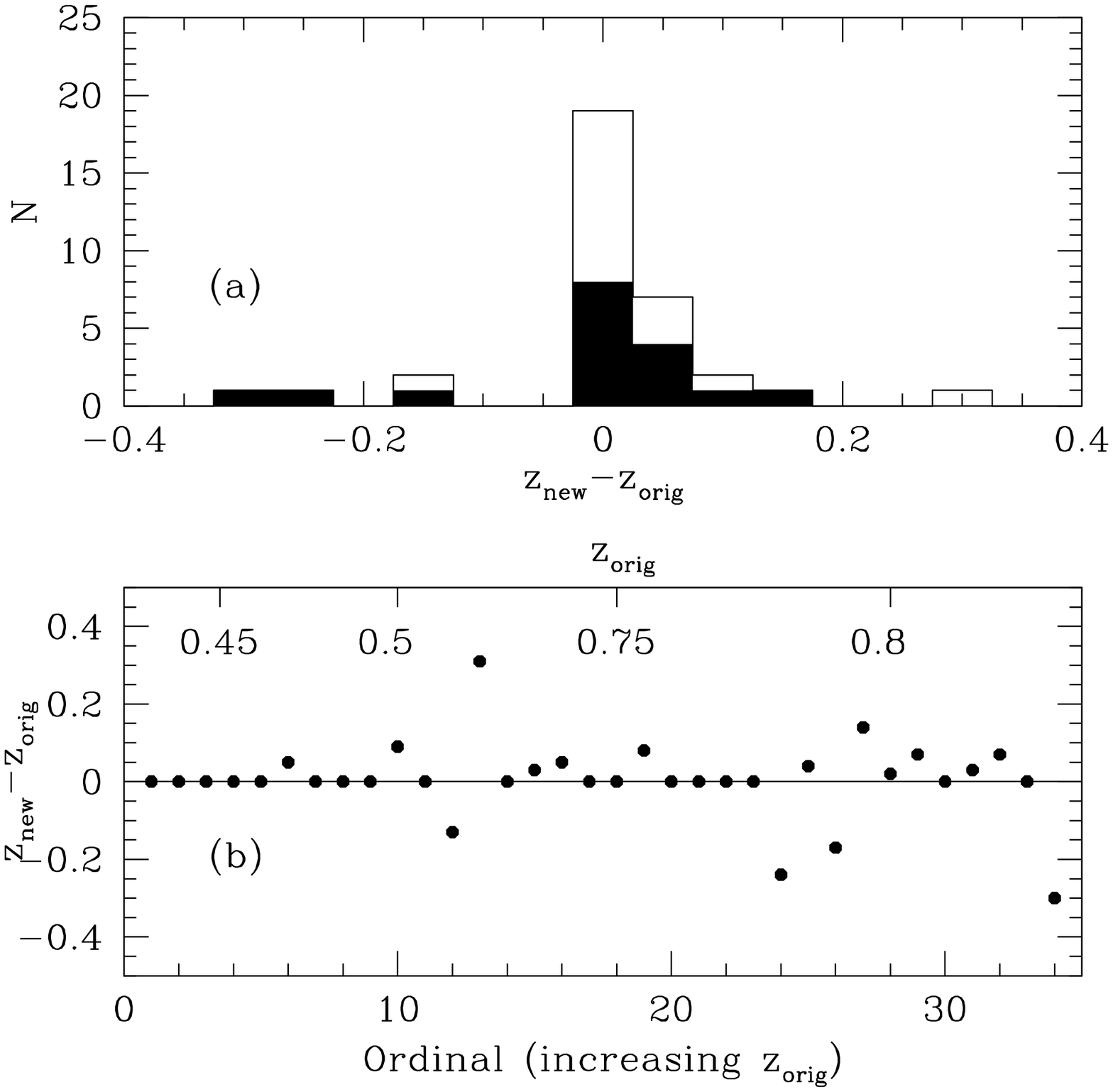}
\figcaption{Impact on the estimated redshifts of misidentification of the
brightest cluster galaxy in the original LCDCS data. $(a)$ Histogram of the 
redshift error due to misidentification for all 34 candidates with confirmed
red galaxy overdensities.
The filled region denotes clusters with published estimated redshifts $z>0.65$. 
$(b)$ Comparison of the revised estimates with the published values for
the 34 confirmed targets.
\label{fig:dzbcg}}
\end{inlinefigure}

\subsection{BCG Identification}

Because the LCDCS had no color information, the identification of the
BCG was confined within a projected separation of 15$\arcsec$, $\la$
80$h^{-1}$ kpc, from the centroid of the detected surface brightness
fluctuation to minimize contamination. An evident concern is that the
BCG may not lie within a projected separation of 15$\arcsec$ and so we
may overestimate the cluster redshift. To test whether this is a
significant problem we have compared our initial results with those
obtained using color-selected BCGs and much larger search radii and
find only a modest effect \citep{n02}. We redo that test with these
new data.

We define the BCG to be the brightest galaxy within a projected radius
of 350 $h^{-1}$ kpc from the peak of the galaxy distribution that has
a color within 0.2 mag of the fitted E/S0-sequence. 
To implement color selection, we obtain a more precise measurement of
the location of the E/S0 sequence by using the adaptive kernel to
perform a second, finer grid search through color space using a
narrower color window ($\pm$0.2 mag) than we did for cluster
identification.  While the EDisCS data is employed for the color
selection, we use the LCDCS survey data in the $W$ passband (the wide,
optical filter used for the LCDCS) to obtain the magnitude of the
brightest galaxy (the LCDCS magnitude-redshift relation is calibrated
in the $W$-band). The search radius is a compromise between maximizing
the probability of including the BCG and minimizing the number of
bright foreground galaxies with similar colors. We choose a radius of
350 $h^{-1}$ kpc ($65-100 \arcsec$ at $z=0.85-0.35$) because 90\% of
local BCG's in Abell and ACO clusters lie within this radius of the
cluster center \citep{post95}.  Because three of the clusters (LCDCS
0632, LCDCS 0633, LCDCS 0634) are at small projected separations from
one another, we decrease the search radius to 200 $h^{-1}$ kpc for
these three systems.  We find that 79\% of BCGs lie within 15$\arcsec$
of the centroid of surface brightness fluctuation detected in the
LCDCS. The concentration of color-selected BCGs at projected
separations $\ll 350 h^{-1}$ kpc (see Figure \ref{fig:bcghist}$b$)
suggests that the typical BCG is within this projected radius at both
$z\simeq0.5$ and $z\simeq0.8$.

\subsection{Estimated Redshifts} \label{sec:zest}

Once the BCG is identified, we use the original LCDCS photometry and
Equation 8 from \citet{gon2001} to determine revised estimated
redshifts.  Figure \ref{fig:dzbcg}$a$ presents a histogram of the
differences in estimated redshifts relative to the original catalog.
For the 34 candidates with confirmed galaxy overdensities from \S
\ref{sec:clusterconf}, the BCG identification differs from
\citet{gon2001} in sixteen cases (47\%).  In 12 of these cases, the
BCG from \citet{gon2001} fails the color selection (including three
objects that are point sources in the deeper imaging). These objects
are typical bluer than the red sequence and are likely foreground
contaminants (although some may indeed be cluster galaxies with strong
star formation). In the other four cases, the new BCG identification
is a galaxy further than 15$\arcsec$ from the published cluster
location.

Despite the high incidence of BCG misidentification, in only six cases
(18\%) does the new identification change the estimated redshift by
$|\Delta z_{est}|>0.1$. This result is consistent with the published
rms uncertainties of 13\% at $z_{est}\simeq0.5$ and 20\% at
$z_{est}\simeq0.8$ for the estimated redshifts, although as noted by
\citet{gon2001} the uncertainties are not symmetric about $z_{est}$.
In four of the six cases where the change in redshift is significant,
$z$ was overestimated because the BCG is at a projected radius that is
greater than the LCDCS search radius ($15\arcsec $).  In the other two
cases a foreground galaxy and a star were mistakenly identified as the
brightest cluster galaxies in the LCDCS data (which is in a single
filter and has an average image quality of 1.5 arcsec).  Figure
\ref{fig:dzbcg}$b$ shows the impact of these errors.  The net result
is that 3 of the candidates with published redshifts $z_{est}\ge0.75$
should have estimated redshifts $z_{est}<0.65$ based upon the new BCG
identification (19\%), while one cluster with published value
$z_{est}=0.4-0.6$ should have an estimated redshift $z_{est}>0.7$. The
other cluster with a significantly underestimated redshift had a
published value
$z_{est}$=0.78, and so remains  
in the high-redshift bin.  A definitive assessment of the robustness
of the redshift estimates awaits spectroscopic observations of the
EDisCS candidates, but this analysis indicates that targeted searches
using the LCDCS catalog should expect of order 15-20\% of candidates
to have redshifts that differ from that published by $\Delta z>0.1$
due to BCG misidentification. This conclusion agrees with that of
\citet{gon2001}.
\begin{inlinefigure}
\plotone{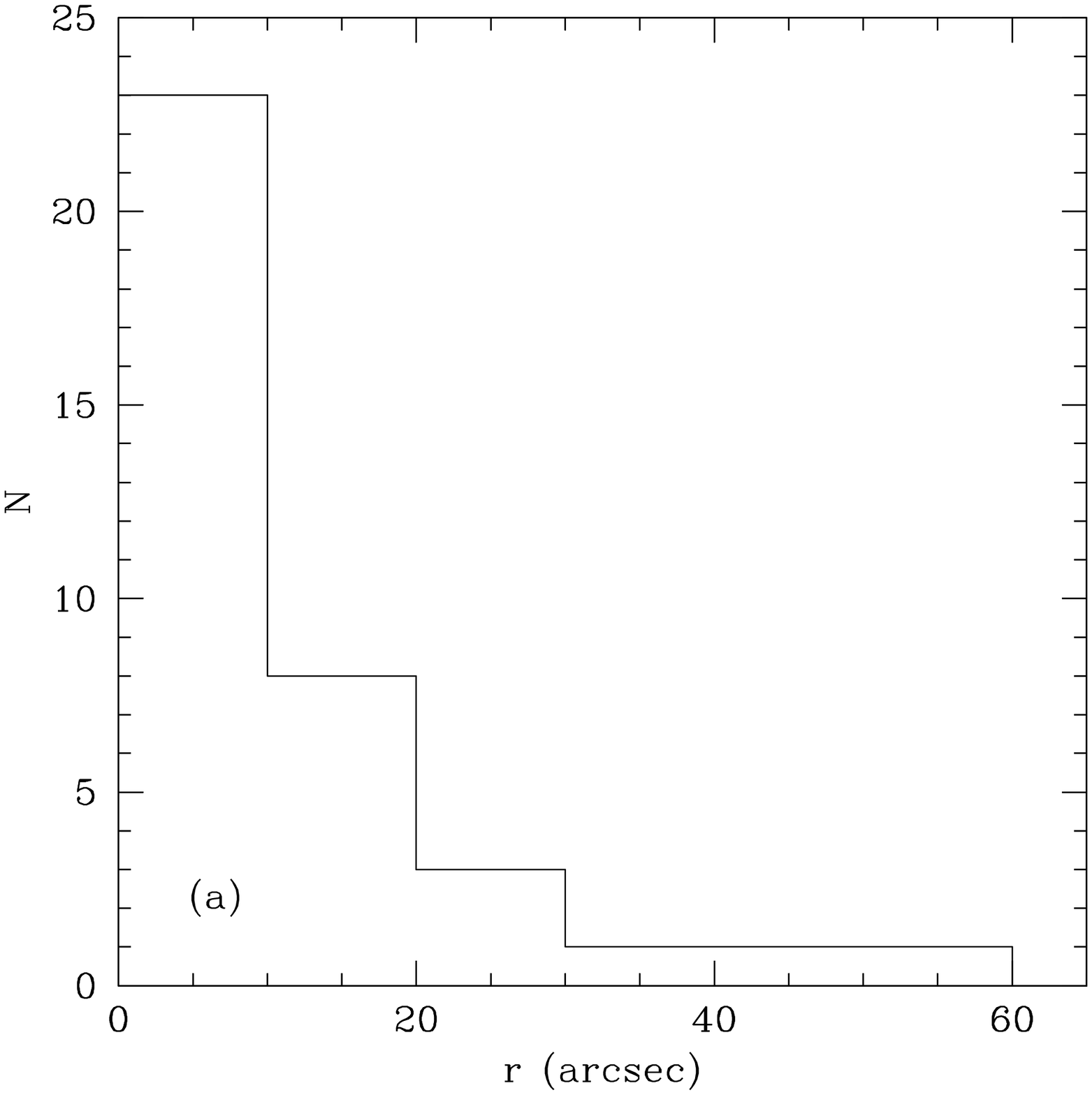}\\
\plotone{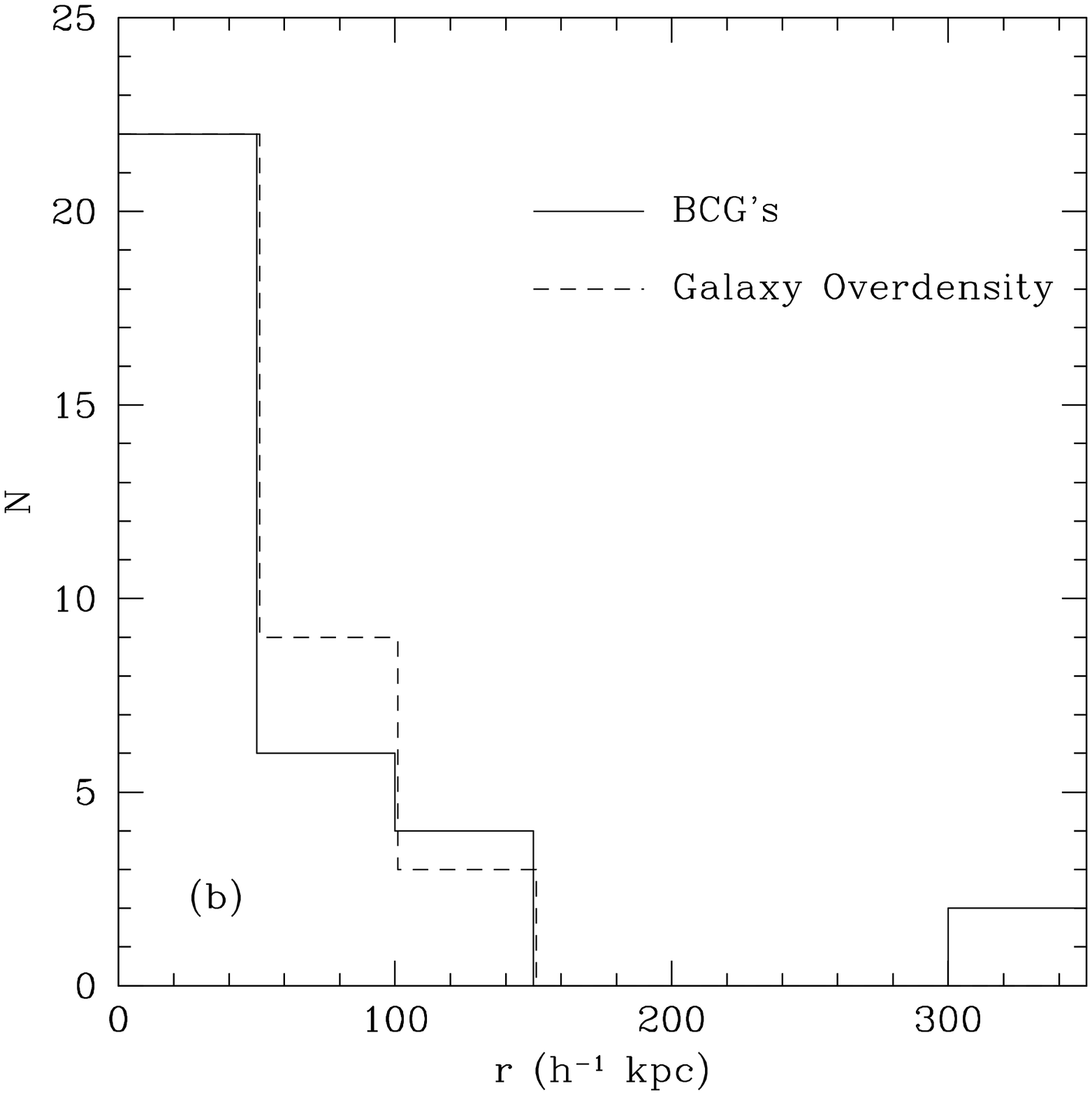}
\figcaption{
$(a)$ - The distribution of angular separations between the published LCDCS 
coordinates and the peak galaxy overdensity for the 37 candidates with
3$\sigma$ detections within 1$\arcmin$ of the published coordinates.
$(b)$ - The solid histogram is the distribution of projected separations (in $h^{-1}$ kpc) between the LCDCS coordinates and the BCG for the 34 candidates with
confirmed red galaxy overdensities
from \S \ref{sec:clusterconf}. For comparison, the overlaid dashed 
histogram shows the separations between the LCDCS coordinates and peak
galaxy density for these clusters.
\label{fig:bcghist}}
\end{inlinefigure}

\subsection{Alignments and Structure}

The LCDCS detection of ``clustered" photons does not discriminate
amongst the possible sources of those photons. The surface brightness
feature could be dominated by unresolved galaxies or the extended BCG
halo. The degree to which each of those contributes to the surface
brightness fluctuation dictates what type of cluster we are most
likely to identify and impacts the estimate of the cluster mass.
We compare the alignment of the BCGs and peak galaxy overdensities
with
\begin{inlinefigure}
\plotone{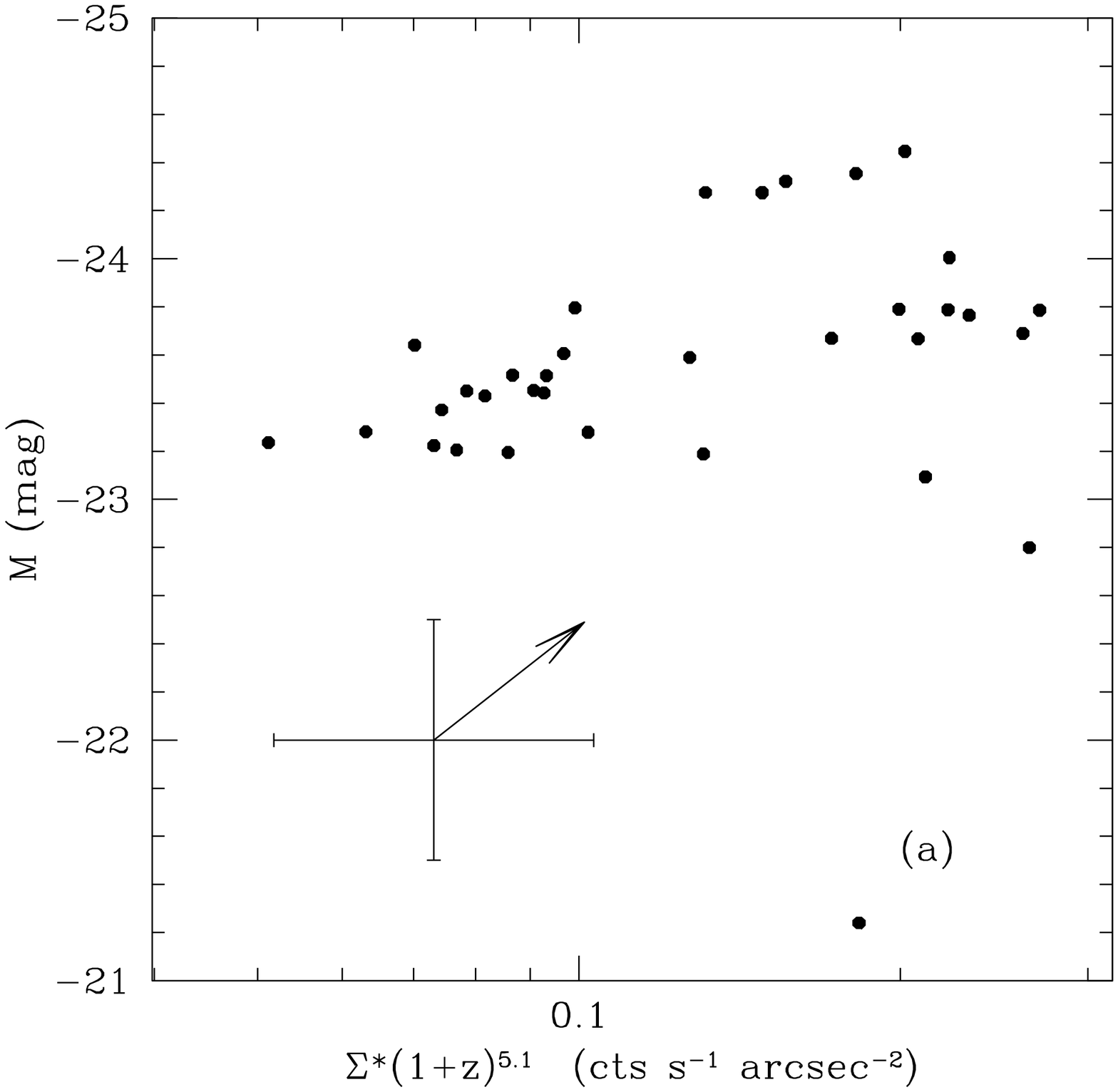}\\
\plotone{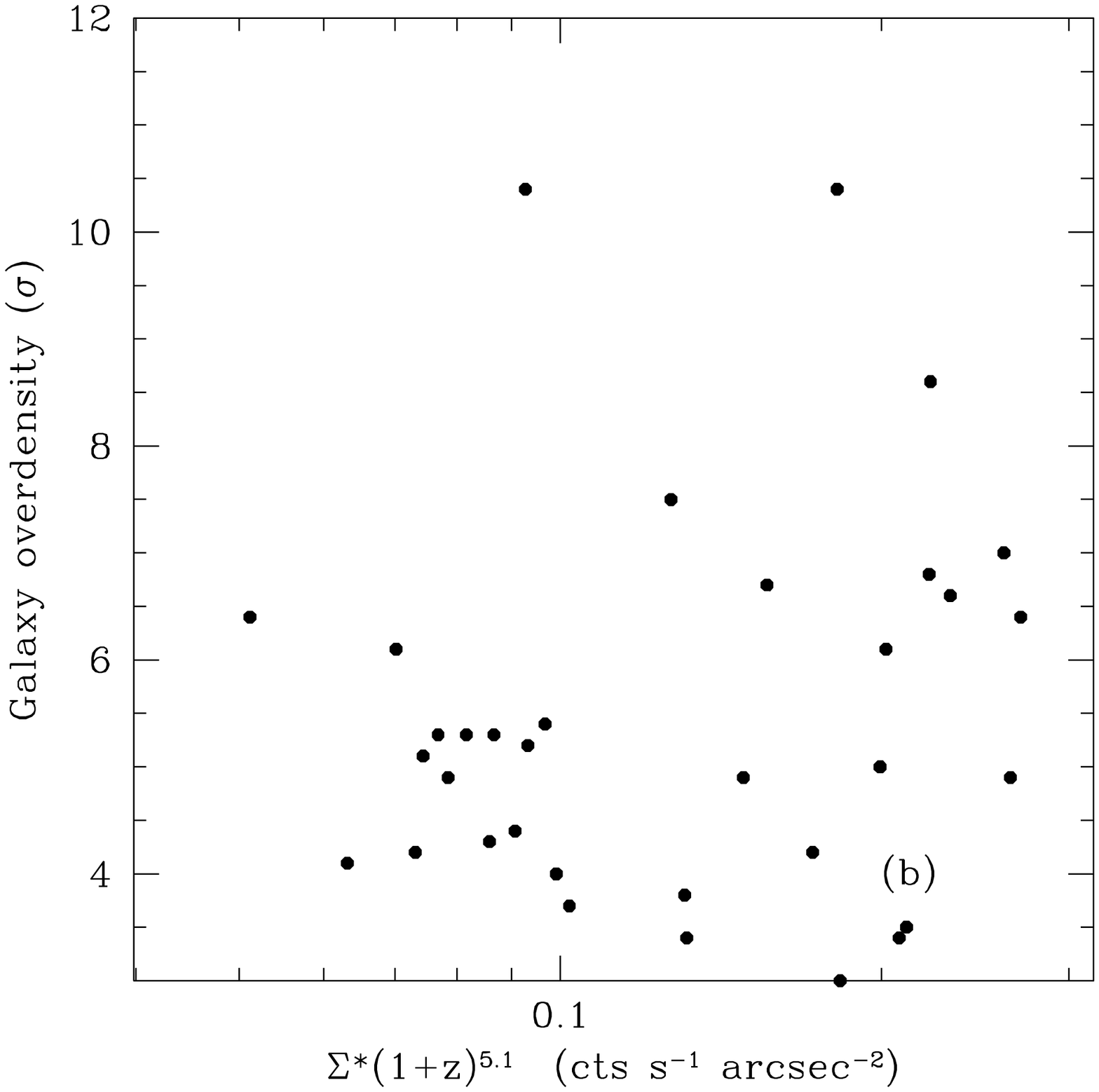}
\figcaption{(a) Absolute magnitude of the red sequences as a function of
surface brightness, $\Sigma$. The surface brightness has been multiplied by
$(1+z)^{5.1}$ to account for E+$k$ corrections and cosmological dimming \citep{gon2000}.
The cross in the lower left corner denotes the typical error bars for each data point
(including redshift uncertainty), while the arrow shows the magnitude and direction
by which a data point would move if the redshift estimate were increased by 0.1.
(b) Significance of galaxy overdensity as a function of $\Sigma$.
\label{fig:sbinfo}
}
\end{inlinefigure}

\noindent the LCDCS detections in Figure \ref{fig:bcghist}.  It appears
that these clusters have established their dynamical centers and that
galaxies within the core region have relaxed into a common potential
well by $z \sim 0.8$.  While the brightest cluster galaxy locations
are well-correlated with the LCDCS detections (79\% lie within
15$\arcsec$), we note that they are no better aligned than the peak
galaxy overdensity. For the six cases in which the peak galaxy
overdensity and BCG location are not coincident ($>15 \arcsec$
separation), the LCDCS coordinates are closer to those of the galaxy
overdensity than to those of the BCG in four instances. Thus, the
LCDCS surface brightness detections appear to be more sensitive to the
contribution of the unresolved cluster galaxies than to the light from
the extended halo of the BCG.

\section{The LCDCS Mass Estimates}
In addition to redshift estimates, the LCDCS also provides a coarse
means of estimating cluster masses. Specifically, \citet{gon2001}
finds that the peak surface brightness of the LCDCS cluster
detections, $\Sigma_{cor}$, is strongly correlated with X-ray
temperature and velocity dispersion (linear correlation coefficients
$r$=0.80 and $r$=0.82) for a calibration sample of roughly ten
clusters.  While we are unable to test these mass predictions with the
current data set, we do check whether the peak surface brightness is
correlated with either the absolute magnitude of the cluster red
sequence (Figure \ref{fig:sbinfo}$a$) or the significance of the
galaxy overdensity detection (Figure \ref{fig:sbinfo}$b$).  For the
former, we compute the total background-subtracted magnitude for
galaxies within 100$h^{-1}$kpc of the peak galaxy overdensity with
colors within 0.2 mag of the red sequence. We then compute the
absolute magnitude for a flat $\Omega_0=0.3$ $\Lambda$CDM cosmology.
We find that there is a weak correlation between the absolute
magnitude of the red sequence and the peak surface brightness
($r$=0.4); however, we refrain from deriving quantitative conclusions
at present. Redshift errors move points on a track nearly parallel to
the observed correlation (as indicated by the arrow in the Figure),
and are sufficiently large to preclude robust conclusions.
Spectroscopic redshifts will be required to disentangle this effect.
We also find that the significance of the galaxy overdensity peak is
weakly correlated with the absolute magnitude of the red sequence
($r$=0.4).  However, we find no significant correlation between the
strength of the galaxy overdensity and the LCDCS surface brightness
(Figure \ref{fig:sbinfo}$b$), possibly due to the large scatter in
both quantities.

\section{Summary and Discussion}
In this paper we present results from the initial phase of the ESO
Distant Cluster Survey (EDisCS), using VLT imaging to better
characterize the Las Campanas Distant Cluster Survey catalog. We first
use smooth density maps of the color-selected galaxy distribution to
confirm cluster candidates. We find that 93\% (28/30) of the EDisCS
targets are coincident with statistically significant overdensities of
red galaxies, as are 60\% (6/10) of serendipitously imaged LCDCS
candidates. The latter number is consistent with the contamination
rate published by \citet{gon2001} for a randomly-selected subsample of
the LCDCS with the same redshift distribution. In addition to
confirming a set of promising LCDCS candidates for further study, we
also use the photometry to identify the brightest cluster galaxies and
use these identifications to test the robustness of the estimated
redshifts published in the LCDCS catalog.  We find that
misidentification leads to redshift errors $\Delta z>0.1$ in 6/34
cases (18\%), which is also consistent with predictions from
\citet{gon2001}.  In addition, we find that the surface brightness
detection technique appears to be slightly more sensitive to the
overdensity of unresolved cluster galaxies than it is to diffuse
emission from the extended halos of brightest cluster galaxies,
indicating that the redshifts are most likely to be overestimated for
dynamically unrelaxed systems in which the BCG and peak galaxy
overdensity are not aligned.

Of more general interest are our findings that 1) the distribution of
red cluster galaxies is generally regular and highly centrally
concentrated out to $z \sim 0.8$, and 2) that the BCGs are also found
near the concentrations of red galaxies, suggesting that the cores of
clusters out to $z \sim 0.8$ are typically dynamically relaxed. The
use of photometric and spectroscopic redshifts obtained as part of
EDisCS will help establish whether these conclusions hold once cluster
members are identified.
The above results verify that the surface brightness fluctuation
technique proposed by \citet{dal1995} and employed by \citet{gon2001}
is an effective method of identifying distant clusters, and
demonstrate that the utility of the LCDCS catalog redshifts is not
seriously compromised by misidentification of the BCG. Upcoming
spectroscopy for the EDisCS will improve upon this analysis by
directly testing the robustness of the redshift estimates, as well as
the LCDCS predictions for the cluster velocity dispersions. These data
define the sample of clusters that will comprise the EDisCS.

\section{Acknowledgments}
The authors thank the anonymous referee for suggestions that improved
this manuscript. DZ acknowledges a fellowship from the David and
Lucile Packard Foundation.

\clearpage

\begin{deluxetable}{lrrllrrl}
\tablewidth{0pt}
\tablecaption{Cluster Confirmation}
\tiny
\tablehead{
\multicolumn{4}{l}{Primary EDisCS Targets} & \multicolumn{4}{l}{Serendipitously Observed Candidates}\\
\hline\hline
\colhead{Candidate} &\colhead{$z_{est}$}&  \colhead{$\sigma$} & \colhead{Comments} &
\colhead{Candidate} &\colhead{$z_{est}$}&  \colhead{$\sigma$} & \colhead{Comments} 
}
\startdata
LCDCS 0057      & 0.47 &        4.9 &    & & & & \\
LCDCS 0110      & 0.78 &        5.4 & $\Delta z$=$-0.24$ & LCDCS 0109  & 0.44 &    2.8 &   \\  
LCDCS 0130      & 0.80 &        6.4 &   & LCDCS 0127  & 0.38 &    4.1 & ($>$25$\arcsec$) \\
LCDCS 0172      & 0.76 &        4.9 &     & & & & \\
LCDCS 0173      & 0.83 &        6.8 &    & & & & \\
LCDCS 0188      & 0.52 &        6.4 & $\Delta z$=$-0.13$  & LCDCS 0190  & 0.60 &    4.9 & ($>$25$\arcsec$)   \\ 
LCDCS 0198      & 0.83 &        6.6 &    & & & & \\
LCDCS 0237      & 0.78 &        3.7 & $\Delta z$=$-0.17$   & & & & \\
LCDCS 0252      & 0.63 &        7.5 &    & & & & \\
LCDCS 0275      & 0.77 &        3.4 &    & & & & \\
LCDCS 0294      & 0.75 &        7.0 &    & & & & \\
LCDCS 0340      & 0.85 &        5.3 & $\Delta z$=$-0.30$  & LCDCS 0337  & 0.47 &    4.1 &   \\ 
LCDCS 0430      & 0.49 &        4.2 & & & & & \\
LCDCS 0458      & 0.79 &        3.9 & ($>$25$\arcsec$) & LCDCS 0457  & 0.74 &    6.7 &   \\ 
LCDCS 0504      & 0.80 &       10.4 &    & & & & \\
LCDCS 0531      & 0.79 &        6.1 &    & & & & \\
LCDCS 0541      & 0.48 &        5.3 &    & & & & \\
LCDCS 0567      & 0.47 &        2.9 &    & & & & \\
LCDCS 0634      & 0.48 &        6.0 &   & LCDCS 0632  & 0.57 &    4.0 &   \\
                &      &            &   & LCDCS 0633  & 0.43 &    5.1 &   \\ 
LCDCS 0642      & 0.54 &        4.3 &   & LCDCS 0641  & 0.66 &    3.8 &   \\ 
LCDCS 0665      & 0.78 &        3.5 &    & & & & \\
LCDCS 0674      & 0.78 &        4.2 &    & & & & \\
LCDCS 0713      & 0.53 &        5.0 & $\Delta z$=$0.31$    & & & & \\
LCDCS 0849      & 0.50 &        4.4 &    & & & & \\
LCDCS 0853      & 0.84 &        8.6 &    & & & & \\
LCDCS 0855      & 0.51 &        5.3 &   & LCDCS 0854  & 0.43 &    3.4 &   \\
LCDCS 0925      & 0.47 &       10.4 &   & LCDCS 0926  & 0.68 &    1.1 & Spurious (LSB)   \\ 
LCDCS 0952      & 0.49 &        5.2 &    & & & & \\
LCDCS 0975      & 0.78 &        4.9 & $\Delta z$=$0.14$   & & & & \\
LCDCS 1027      & 0.77 &        3.0 &    & & & & 
\\
\enddata
\label{tab:confirmation}
\tablecomments{In the comments, $>$25$\arcsec$ refers to the separation between the peak
galaxy overdensity and the published LCDCS coordinates, while $\Delta z\equiv z_{new}-z_{est}$
denotes the change in estimated redshift due to the revised identification of the BCG. We
only list this value for candidates where the redshift estimate changed by $\Delta z>0.1$.
}
\end{deluxetable}


\begin{thebibliography}{}
\bibitem[Aragon-Salamanca, Baugh, \& Kauffmann(1998)]{ara1998}  Aragon-Salamanca,
A., Baugh, C. M., Kauffmann, G. 1998, \mnras, 297, 427
\bibitem[Aragon-Salamanca et al(1993)]{ara1993}  Aragon-Salamanca, A.,
Ellis, R. S., Couch, W. J., Carter, D. 1993, \mnras, 262, 764   
\bibitem[Bertin \& Arnouts(1996)]{ber96} Bertin, E., \& Arnouts, S. 1996, \aaps,
313, 21
\bibitem[Bruzual \& Charlot(1993)]{bru93} Bruzual, A., Charlot, S. 1993, \apj, 405, 538
\bibitem[Burke, Collins, \& Mann (2000)]{burke2000}Burke, D.J., Collins, C.A., \& Mann, R.G.
2000, \apj,  532, L105
\bibitem[Charlot, Worthey, \& Bressnan(1996)]{cha96} Charlot, S., Worthey, G., \& Bressnan, A. 1996, \apj, 4 57, 625
\bibitem[Collins \& Mann(1998)]{col1998} Collins, C. A., \& Mann, R. G. 1998,
\mnras, 297, 128 

\bibitem[Dalcanton(1995,1996)]{dal1995} Dalcanton, J. J. 1995, Ph. D. thesis
\bibitem[Dalcanton(1996)]{dal1996} Dalcanton, J. J. 1996, \apj, 466, 9
\bibitem[Gladders \& Yee(2000)]{glad2000} Gladders, M.D., \& Yee, H.K.C. 2000, \aj, 120, 214
\bibitem[Gonzalez(2000$a$)]{thesis} Gonzalez, A. H. 2000, Ph. D. Thesis,
University of California at Santa Cruz.
\bibitem[Gonzalez et al.(2000$b$)]{gon2000} Gonzalez, A. H., Zabludoff, A.I., 
Zaritsky, D., \& Dalcanton, J. J. 2000, \apj, 536, 561
\bibitem[Gonzalez et al.(2001)]{gon2001} Gonzalez, A. H., Zaritsky, D.,
Dalcanton, J. J., \& Nelson, A. E. 2001, \apjs, 137, 117
\bibitem[Gonzalez, Zaritsky, \& Wechsler (2002)]{gon2002} Gonzalez, A.H., Zaritsky, D., \& Wechsler, R.H. 2002, 
\apj, in press
\bibitem[Merritt \& Tremblay(1994)]{mer94} Merritt, D. \& Tremblay, B. 1994, \aj,
108, 514
\bibitem[Nelson et al.(2001)]{nelson2001}Nelson, A.E., Zaritsky, D., Gonzalez,
A.H., \& Dalcanton, J.J. 2001, \apj, 563, 629
\bibitem[Nelson et al. (2002)]{n02}Nelson, A.E., Gonzalez, A.H., Zaritsky, D., and \& Dalcanton, J.J. 2002, \apj, 566, 103
\bibitem[Olsen et al.(2001)]{olsen2001} Olsen, L. F. et al. 2001, \aap, 380, 460
\bibitem[Postman \& Lauer(1995)]{post95}Postman, M., \& Lauer, T.R., 1995,
\apj, 440, 28
\bibitem[Postman et al.(1996)]{post96}Postman, M., et al. 1996, \aj, 111, 615
\bibitem[Sandage(1988)]{san1988} Sandage, A. 1988, \araa, 26, 561 
\bibitem[Stanford, Eisenhardt, \& Dickinson(1998)]{stanford1998}Stanford, A.,
Eisenhardt, P., \& Dickinson, M. 1998, \apj, 492, 461
\bibitem[White et al. (2002)]{white02}White, S. D. M., et al. 2002, in prep.
\bibitem[Zaritsky et al.(1997)]{zar1997} Zaritsky, D., Nelson, A. E., Dalcanton,
J. J., \& Gonzalez, A. H.  1997, \apj, 480, L91  
\end{thebibliography}
\end{document}